\documentclass[journal]{IEEEtran}

\usepackage{amsmath}
\usepackage{algorithm}
\usepackage[noend]{algpseudocode}
\algnewcommand\algorithmicforeach{\textbf{for each}}
\algdef{S}[FOR]{ForEach}[1]{\algorithmicforeach\ #1\ \algorithmicdo}
\usepackage{tabularx}
\ifdefined\usePortuguese
\usepackage[brazil]{babel}   
\usepackage[utf8]{inputenc}
\else
\fi

\ifdefined\appendices
\else
\ifdefined\addAppendixAsPreamble 
\usepackage[titletoc,title]{appendix} 
\else
\usepackage{appendix} 
\fi
\fi

\usepackage{makeidx}  
\usepackage{graphicx}
\usepackage{color} 
\definecolor{darkgreen}{rgb}{0,0.35,0}

\ifdefined\keywords
\else

	\def\keywords{\vspace{.5em}{\bfseries\textit{Index Terms}---\,\relax%
	}}
	\def\endkeywords{\par}
\fi

\usepackage{url} 

\ifdefined\RCSfile 
\else
\usepackage{cite} 
\fi

\usepackage{multirow} 

\usepackage[bbgreekl]{mathbbol}

\ifdefined\amsmath
\else
\usepackage{amsmath}
\fi

\ifdefined\amsfonts
\else
\usepackage{amsfonts}
\fi

\usepackage{amssymb}

\ifdefined\proof
\else
\usepackage{amsthm}
\fi

\ifdefined
\else
\ifdefined\akbook
\ifdefined\lulu
\typearea[6mm]{1} 
\else
\usepackage[a4paper, top=3cm, bottom=3.4cm, left=1.8cm, right=2.5cm]{geometry}
\fi
\fi
\fi

\usepackage{mathrsfs} 
\usepackage{makeidx}  
\usepackage{graphicx,color}

\usepackage{multirow} 
\usepackage{listings} 
\usepackage{ifpdf} 

\ifpdf

\ifdefined\hyperref
\else
\usepackage[pdftex,colorlinks]{hyperref}
\fi

	\usepackage[pdftex]{thumbpdf} 
	\usepackage{pdflscape}
\usepackage{epstopdf}
\hypersetup{%
pdftitle={Some title},
pdfauthor={Your name - LaPS - UFPA},
pdfkeywords={DSP,Signal},
pdfstartview={FitH}, 
urlcolor=black,
linkcolor=black,
citecolor=black,
}

\fi 

\ifdefined\bm
\else
\newcommand{\bm}[1]{{\mbox{\boldmath $#1$}}}
\fi

\ifdefined\Theorem
\else
\newtheorem{Theorem}{Theorem}
\fi

\newtheorem{Theorem*}{Theorem}

\newtheorem{Application*}{Application}

\newtheorem{Claim*}{Claim}

\ifdefined\Corollary
\else
\newtheorem{Corollary}[Theorem]{Corollary}
\fi

\newtheorem{CounterExample*}{$\overline{\hbox{\bf Example}}$}

\ifdefined\Example
\else
\newtheorem{Example}{Example}
\fi

\newtheorem{Example*}{Example}

\newtheorem{Intuition*}{Intuition}
\newtheorem{Joke*}{Joke}

\ifdefined\Lemma
\else
\newtheorem{Lemma}[Theorem]{Lemma}
\fi

\newtheorem{Lemma*}{Lemma}

\newtheorem{Model*}{Model}
\newtheorem{Open problem}{Open problem}

\newtheorem{Question*}{Question}
\newtheorem{Remark*}{Remark}

\def \bSubexa    {\begin{subexa}}

\ifdefined\Problem
\else
\newenvironment{Problem}{\textbf{Problem}\\ \begin{enumerate}}{\end{enumerate}}
\fi

\newcommand{\ignore}[1]{{}}

\def \bf {{\textbf f}}

\def \bm {{\textbf m}}

\def \mynote#1{{}}

\def\ignore#1{}

\def\orpro{\mathop{\mathchoice
   {\vee\kern-.49em\raise.7ex\hbox{$\cdot$}\kern.4em}
   {\vee\kern-.45em\raise.63ex\hbox{$\cdot$}\kern.2em}
   {\vee\kern-.4em\raise.3ex\hbox{$\cdot$}\kern.1em}
   {\vee\kern-.35em\raise2.2ex\hbox{$\cdot$}\kern.1em}}\limits}

\def\andpro{\mathop{\mathchoice
 {\wedge\kern-.46em\lower.69ex\hbox{$\cdot$}\kern.3em}
 {\wedge\kern-.46em\lower.58ex\hbox{$\cdot$}\kern.25em}
 {\wedge\kern-.38em\lower.5ex\hbox{$\cdot$}\kern.1em}
 {\wedge\kern-.3em\lower.5ex\hbox{$\cdot$}\kern.1em}}\limits}

\usepackage[tight,footnotesize]{subfigure}

\newcommand{\mytitle}{Downlink Fronthaul Compression in Frequency Domain using
OpenAirInterface}

\begin{document}

\title{\mytitle}

\author{\IEEEauthorblockN{Cleverson Nahum, Leonardo Ramalho, Joary Fortuna,
Chenguang Lu, Miguel Berg, \\Igor Almeida and Aldebaro Klautau} \thanks{This
work was supported in part by the Innovation Center, Ericsson
Telecomunica\c{c}\~oes S.A., Brazil, CNPq and the Capes Foundation, Ministry of
Education of Brazil, and by the European Union through the 5G-Crosshaul project
(H2020-ICT-2014/671598).} \thanks{C. Nahum, L. Ramalho, J. Fortuna and A.
Klautau are with LASSE - 5G Group, Av Perimetral km 01, Guam\'a, 66075-750,
Federal University of Par\'a, Bel\'em, Brazil (e-mails: \{cleversonahum,
leonardolr, joary, aldebaro\}@ufpa.br).} \thanks{Chenguang Lu and Miguel Berg
are with Ericsson Research, 164 80, Stockholm, Sweden (e-mails:
{chenguang.lu,miguel.berg}@ericsson.com, ).} \thanks{Igor Almeida is with
Ericsson Research, Rod. Eng. Ermênio de Oliveira Penteado, 57,5 - Parque São
Lourenço, Indaiatuba, Brazil (e-mail: igor.almeida@ericsson.com)}}

\maketitle
\begin{abstract}

This paper presents a compression scheme developed for the transport of downlink
radio signals in packet fronthaul of centralized-radio access networks (C-RAN).
The technique is tailored to frequency-domain functional splits, in which
inactive LTE (or 5G NR) resource elements may not be transmitted in the packets.
This allows decreasing the link data rate, especially when the cell load is low.
The compression scheme is based on two parts: transmission of side information
to indicate active and inactive resource elements of the LTE, and nonuniform
scalar quantization to compress the QAM symbols of the active resource elements.
The method was implemented in the OpenAirInterface (OAI) software for real-time
evaluation. The testbed results show a significant reduction in link usage with
a low computational cost.

\end{abstract}


\section{Introduction}

The fifth-generation (5G) of mobile communications is expected to support a
massive amount of users and provide very high data rates~\cite{dahlman20145g}.
For that, it is required to increase system capacity, reduce communication
latency, and increase reliability. Therefore, some features such as network
densification, use of higher frequencies, and multiple antenna techniques are
being explored~\cite{chen2014requirements}. To support the increased data rates,
flexibility, new capabilities, and other demands of 5G, the Radio Access Network
(RAN) is evolving. An example of this evolution is the Centralized RAN (C-RAN)
architecture~\cite{chih2014recent}, which allows providing cost-effective mobile
network implementation. In C-RAN an eNodeB is split into Remote Radio Unit (RRU)
and centralized Baseband Unit (BBU), which communicate over a fronthaul (FH)
link. BBUs are allocated in the data center, where it is possible to share
computational resources amongst different sites providing more flexibility and
costs saving. A drawback of C-RAN is the required high bit rate of the FH, which
is the link used to transport the radio signal between BBU and
RRU~\cite{bartelt2015fronthaul}. 

There are different approaches to decrease the FH rate in the literature, such
as time-domain signal compression and functional split~\cite{larsen2018survey}.
For example, time-domain signal compression can be based on well-known methods
such as vector quantization~\cite{Si17} or linear prediction and
Huffman~\cite{ramalho2017lpc}. On the other hand, with functional split, the
network processing stack is split between the BBU and RRU. Splits allow to
decrease the FH bit rate, but the RRU performs more computational processing.
Another solution is to combine both and use signal compression
algorithms~\cite{park2014fronthaul,pawar2015front} along with functional splits.
An example of FH compression with the functional split is given in
\cite{lorca2013lossless}, where the authors propose a Baseband Fronthaul
Termination (BFT) Nodes to make compression/decompression on FH and allocate
resource blocks according to the cell usage reaching a fronthaul reduction of
33,3\% in a full load system. One drawback of~\cite{lorca2013lossless} is the
huge increase of the computational complexity in the RRU because of a large part
of the PHY layer is implemented in the RRU. A flexible functional split with
A-law lossy compression for FH signals was presented in \cite{chang2017flexcran}
reaching a fronthaul reduction of 50\%. The Low-PHY split 7.1 was implemented in
\cite{chang2017flexcran}, which reduced the RRU computational cost but increased
the FH bit rate because all subcarriers of an LTE symbol are sent, regardless of
carrying or not useful data.


\begin{figure*}[!t]
\begin{center}
 \includegraphics[width=0.85\linewidth]{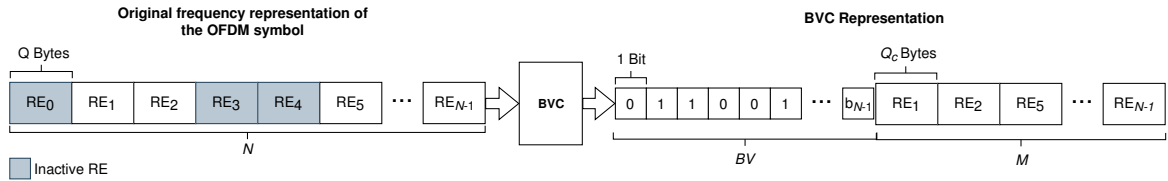}%
 \caption{Example of the bit vector compression (BVC) applied to a OFDM symbol,
 where each RE is as QAM symbol that modulates a subcarrier.
 \label{fig:bvc}}
\end{center}
\end{figure*}

This paper presents an algorithm to compress and transport frequency domain
representation of the LTE OFDM symbols. The proposed method is a combination of
functional split and signal processing, where the RRU is responsible for two
parts of the baseband OFDM signal generation: the IFFT and cyclic prefix
insertion. The proposed algorithm, called \emph{Bit Vector Compression} (BVC),
discards the unused subcarriers before transporting data over the FH. Thus, the
bit rate of the FH is proportional to the number of active subcarriers. After
discarding the unused subcarriers, the IQ samples are quantized to reduce the FH
bit rate further. In this paper, we used an A-law quantizer, but other scalar
and vector quantizers can be used. We implemented the BVC in the Eurecom's Open
Air Interface (OAI) software~\cite{oai} to evaluate it in an online network. As
in the original OAI's implementation, the LTE OFDM symbols are transported with
the frequency domain representation over UDP packets in an Ethernet
FH~\cite{chang2017flexcran}. Since unused LTE Resource Elements (RE) are removed
in this paper before the packet is sent through the FH, it is possible to reduce
the FH data rate according to the data rate requested by the User Equipments
(UEs). We show real-time results of data rate reduction and the relation between
the FH and UEs' data rates.


Some of the contributions of the paper are listed in the sequel. The proposed
algorithm has low complexity and can be used in scenarios of low-PHY functional
split 7.1~\cite{3gpp2017study} to send only used subcarriers from LTE symbols.
The proposed compression method allows decreasing the FH rate according to the
UEs' data rates. Furthermore, the proposed method could be implemented in
low-cost RRUs, since the selected functional split implements few operations of
the signal generation, and the proposed compression method is simple. Lastly,
the results of the proposed algorithm are shown in a real testbed that can be
modified in future works to implement other methods.

This paper is organized as follows. Section~\ref{sec:method} presents a detailed
description of the proposed compression method, explaining the algorithm and
resources estimation accordingly to the number of active subcarriers.
Section~\ref{sec:implementation} shows the BVC implementation using an OAI
testbed. Experimental results from the testbed are presented in
Section~\ref{sec:results}. Section~\ref{sec:conclusion} concludes the paper.

\begin{figure*}[t]
	\centering
	\subfigure[Compression Algorithm flowchart.]{
	\includegraphics[width=0.4\textwidth]{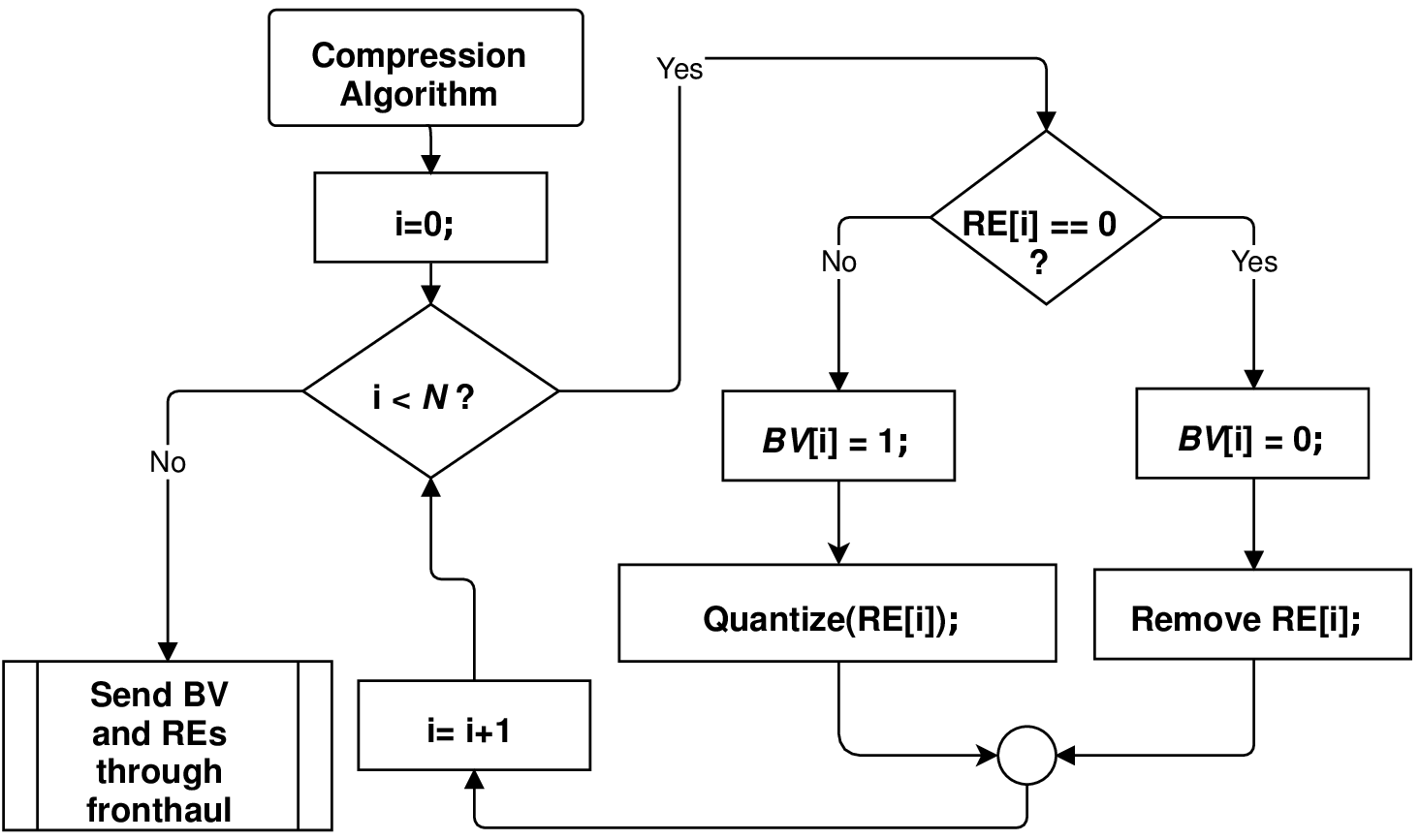}
	\label{fig:CompressionFlowGraph}
	} \subfigure[Decompression Algorithm flowchart.]{
	\includegraphics[width=0.4\textwidth]{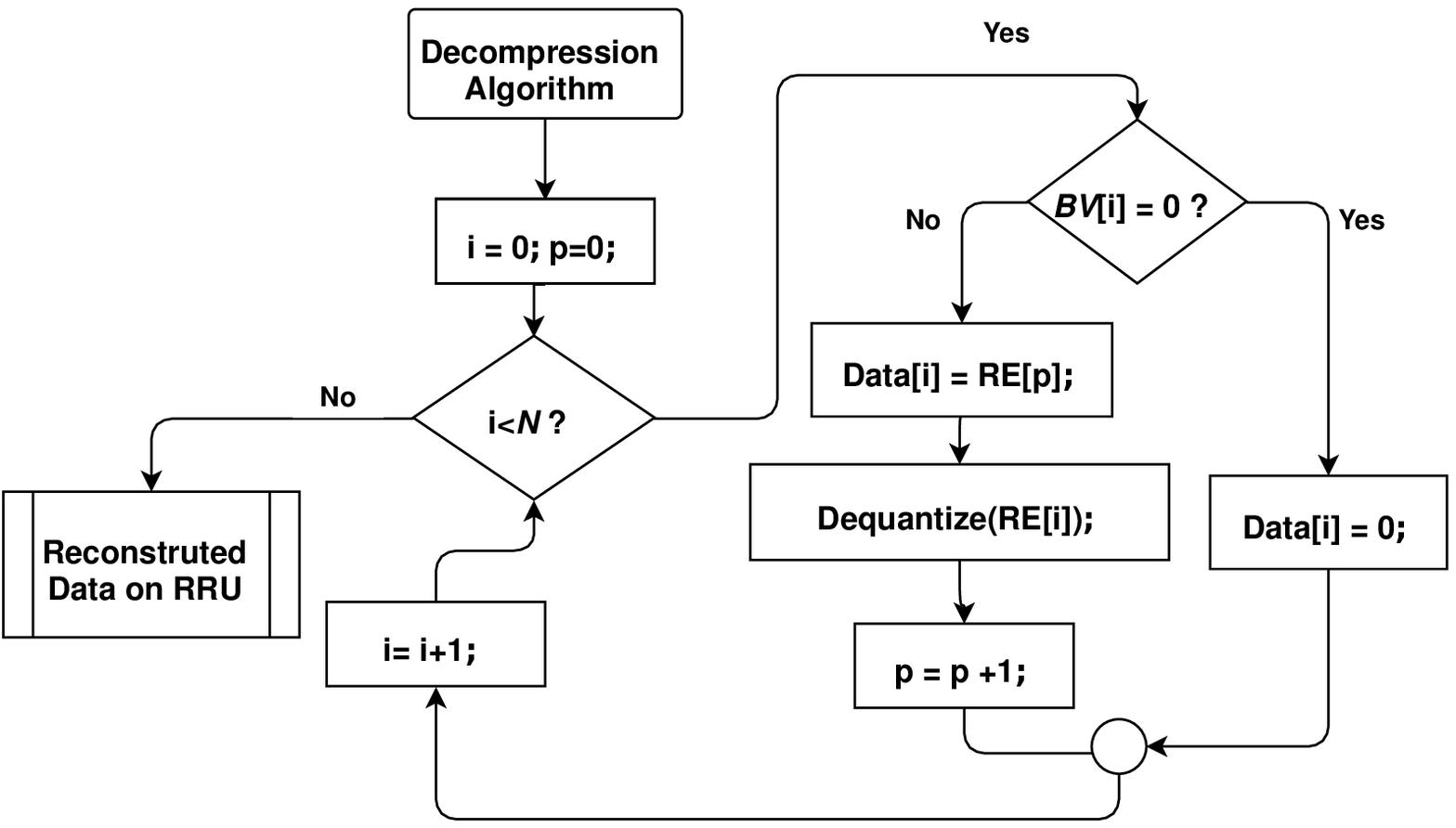}
	\label{fig:DecompressionFlowGraph}
	}
	\caption{Compression and decompression operations of the proposed BVC algorithm.}
	\label{fig:algorithms}
\end{figure*}

\section{Proposed Method}
\label{sec:method}

This section describes the proposed method to transport LTE signals more
efficiently over a packet-based FH. For concreteness, we assume a 4G LTE
network, but the method can also be applied to 5G New Radio (NR) scenarios. More
specifically, we adopt OAI as a \emph{reference implementation} of a
packet-based LTE network scenario using 3GPP's 7.1 Low-PHY
split~\cite{3gpp2017study}, in which frequency domain data (also called
subcarriers or QAM symbols) are sent through FH based on UDP transport. In the
following paragraphs, we first present the foundations of the BVC, and then we
provide equations to calculate the size of the packet that transports the
compressed radio signal.

FH compression methods that work with the time-domain waveform
(e.g.,~\cite{ramalho2017lpc,Si17}) send data corresponding to all REs of an OFDM
symbol, regardless of the RE usage. In contrast, the proposed compression
algorithm removes inactive REs from the packet of the downlink signals to
achieve a bit rate that depends on the load in the wireless interface. It allows
decreasing FH downlink traffic accordingly to the number of active subcarriers,
which is proportional to the UEs data rates. In the BVC, one OFDM symbol is sent
per UDP packet, where active REs and side information are sent. The
side-information is called bit vector (BV) and indicates which subcarriers are
active and inactive. Thus, the receiver can properly reconstruct the OFDM symbol
in the frequency domain. The BV is a sequence of $N$ bits that represents the
activity in the $N$ REs in the OFDM symbol, i.e., a value of 1 indicates active
RE, and 0 indicates unused RE that is not transmitted over the FH.

The BVC is illustrated in Fig.~\ref{fig:bvc}, which also shows the BV field to
indicate the active REs. On left side $N$ REs of an original frequency
representation of the OFDM symbol, where each RE is a complex QAM symbol that is
represented with $Q$ bytes. On the right side, all inactive REs were removed
keeping $M$ active REs, where each RE is quantized using A-law encoder becoming
a representation with $Q_c$ bytes. The bit vector $BV$ is appended to the data
to inform which subcarriers are being transported.

Fig.~\ref{fig:algorithms} shows the compression and decompression operations.
Fig.~\ref{fig:CompressionFlowGraph} shows a flowchart of the operations that BBU
performs before sending the samples to the RRU. The BVC maps the REs into the BV
where each bit $BV[i]$ represents a RE of the OFDM symbol. The algorithm
iterates over the REs verifying unused ones to assigns 0 to the respective BV
position and removing the RE. For the used REs, the $BV[i]$ is assigned one to
represent that the corresponding $RE[i]$ is active and the RE is encoded using
A-law.

The RRU can reconstruct the original frequency domain representation of the OFDM
symbol, after receiving the BV and active subcarriers data. The process is shown
in Fig.~\ref{fig:DecompressionFlowGraph}, which represents the decompression
algorithm. The BV and received REs are stored on $BV$ and $REs$ vectors. The
algorithm iterates over each $BV$ element verifying if it represents an inactive
RE. If the current position in BV is one, then the next IQ sample of the RE
array is decoded using A-law and inserted in the reconstructed array, and if the
BV position is zero a null IQ sample (zero) is added in the reconstructed array.
After all, REs were analyzed and the original data received is recovered to be
used on RRU process.

Considering a FH using UDP transport, where each packet carries one OFDM symbol,
so one LTE sub-frame is sent over 14 packets. The UDP packet size is dependent
on the number of resources elements transported and the IQ symbol quantization.
Using a fixed-length quantization scheme, all REs are represented with the same
size. Besides LTE data, there is the header with information about LTE signal
transported, Ethernet, IPv4, and UDP header. Therefore UDP packet size with no
compression has a fixed value of
\begin{equation}
\label{eqn:pktsize}
P_s = P_h + L_h + N \cdot Q,
\end{equation}
\noindent where $P_h$ represents the Ethernet, IPv4, and UDP header size. $L_h$
represents LTE header with information about the OAI LTE, $N$ is the number of
resource elements and $Q$ is the number of bytes used to represent IQ symbols.
Typical values of $P_h$ and $L_h$ are 42 and 12 bytes, respectively~\cite{oai}.
Using the BVC, the BV field adds an additional overhead proportional to the
number of REs in each UDP packet. Thus, UDP packet size using the BVC can be
calculated as follows:

\begin{equation}
\label{eqn:cpktsize}
P_s = P_h + L_h + \left\lceil\frac{{N}}{8}\right\rceil + M \cdot Q_c,
\end{equation}
\noindent where $M$ represents the number of active resource elements and $Q_c$
represents the number of bytes used in A-law encoder to quantize IQ symbols. 

The compression efficiency varies according to IQ quantization and the number of
REs being used. Considering a quantization of 2 bytes per RE and a 5~MHz
bandwidth (maximum of 300 REs per OFDM symbol), when there are more than 281
active REs, the UDP packets are larger than when no compression is used because
of the 38 bytes BV overhead added for 300 REs as perceived on
Fig.~\ref{fig:frameSize}.

Fig.~\ref{fig:frameSize} shows different packet sizes considering different
cases using BVC and without BVC for a LTE signal of 5~MHz. These values can be
calculated with~(\ref{eqn:pktsize}) and (\ref{eqn:cpktsize}). When BVC is used,
the packet size is evaluated with $Q_c=1$ and $Q_c=2$ bytes per complex sample.
Without BVC, the number of bytes per complex sample is also one and two. When
there is no active RE, the BVC packets have just headers and the bit vector
information. For the cases using BVC, when the packet size is larger than
without method, an additional field could be used to indicate whether the BV
field is present or not. Thus, it could be a way of enabling or disabling online
the BVC to deal with its overhead disadvantage.

\section{Validation on OAI-based Testbed}
\label{sec:implementation}

\subsection{OAI implementation}

The BVC was implemented by modifying the OAI software, which is composed of two
main source code repositories: 'openairinterface5g' (OAI5G) and 'openair-cn'.
The project openair-cn has the source-code implementation of the evolved packet
core (EPC) network. OAI5G repository contains source-code implementation for RAN
LTE stack, it can operate as a complete eNodeB or as BBU and RRU, and also
implement UE. It is also possible to make a functional prototype using Software
Defined Radio (SDR) acting as RRU.

\begin{figure*}[!t]
 \begin{center}
 \includegraphics[width=0.899\linewidth]{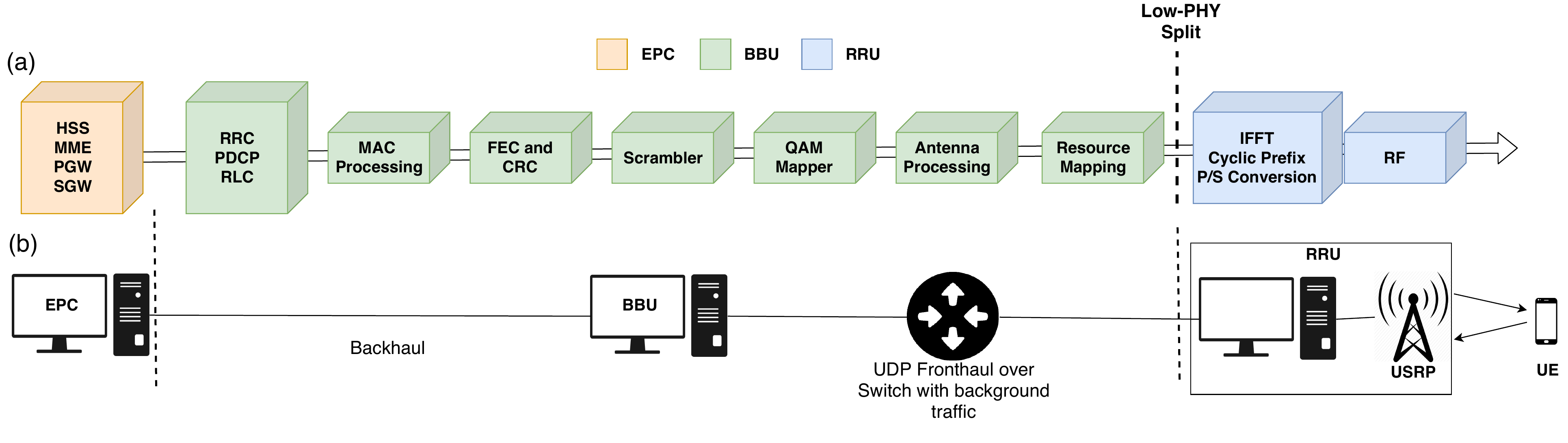}
 \caption{(a) Split 7.1 (Low-PHY) of 3GPP LTE Functional split, which is called
 in OAI as IF4.5. (b) Testbed machines where OAI modules are working and USRP
 used to transmit RF signals to UE (smartphone)}
 \label{fig:testbedSetup}
 \end{center}
 \end{figure*}

Fig.~\ref{fig:testbedSetup} (a) shows the LTE functional split
7.1~\cite{larsen2018survey} defined by 3GPP, which is called in OAI project as
IF4.5 split. In this scenario, RRU implements IFFT, cyclic prefix, parallel to
Serial (P/S) conversion and RF transmission. Fig.~\ref{fig:testbedSetup} (b)
illustrates in which machine OAI modules are executed and their connections. BBU
and RRU are connected through an Ethernet switch, and the packets are UDP
besides background traffic. The Ethernet link is shared with other services
being susceptible to network congestion, delay and packet loss which are handled
by OAI and minimized using the BVC. A USRP B210 is used as an RF transceiver,
and it is connected to the RRU machine through a USB 3.0.

\begin{figure}[thb]
\begin{center}
\includegraphics[width=0.87\columnwidth]{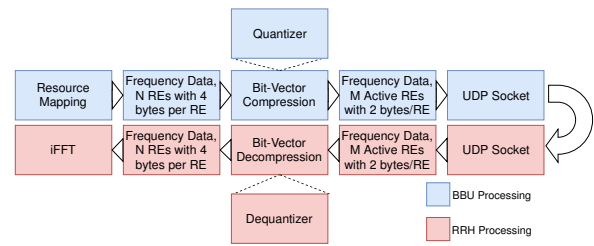}
\caption{Implementation of the bit vector compression and decompression on the
BBU and RRU for downlink signals.\label{fig:process}}
\end{center}
\end{figure}

Fig.~\ref{fig:process} illustrates OAI operations and where the BVC was
implemented. In BBU's process called resource mapping, information is
represented with 4 bytes for each complex-valued RE (2 bytes to the real part
and 2 bytes to imaginary part). BVC removes inactive REs and performs A-law
encoding to the active REs. Then, the information is sent through a UDP socket
and received on RRU, where all inverse operations are applied, and the original
frequency representation is reconstructed. In summary, the OAI software was
altered to include the bit vector compression and bit vector decompression, in
the BBU and RRU, respectively.

\begin{figure}[t]
\begin{center}
 \includegraphics[width=\columnwidth]{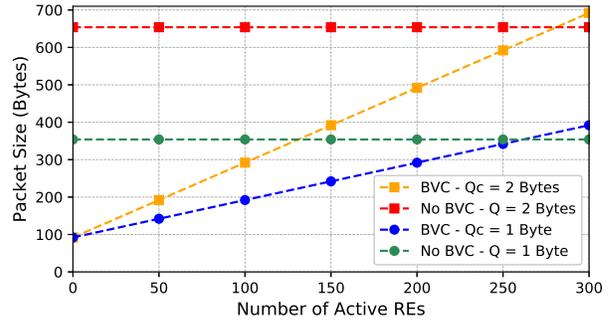}%
 \caption{Variation
 of the packet size with the number of bytes per IQ symbol and number of active
 REs.\label{fig:frameSize}}
\end{center}
\end{figure}

OAI sends one OFDM symbol per UDP packet, so considering a 5~MHz scenario
without compression, it can send 300 REs per packet. Each packet has a header of
$P_h=42$~bytes (UDP, IPv4 and Ethernet headers) plus $L_h=12$~bytes of IF4.5
header added to control frame information and each RE is quantized with
$Q=2$~bytes provided by A-law encoder. So when BVC is not used, to send zero
active REs each packet has 654 bytes, as indicated in Fig.~\ref{fig:frameSize}
because it always sends information, independently if REs are used or not. On
the other hand, when BVC is used, packets with 0 active REs have only 92 bytes
that correspond to header and bit vector information. This value varies
according to the number of active REs.

\subsection{Hardware}

The BBU machine uses a Core i7-5930K@3.50~GHz processor with 16~GB RAM and RRU
machine uses a Core i5-4590@3.30~GHz with 8~GB RAM, both are using kernel
3.19.0-61-lowlatency (real-time operations) and Ubuntu~14.04. EPC runs into a
virtual machine with customized kernel 4.7.7-oaiepc made by OAI inside BBU
machine and using Ubuntu 16.10, sharing resources and with 4~GB RAM allocated.
The OAI was configured to run in Frequency Division Duplex~(FDD) with a central
frequency of 2685~MHz. The LTE signal has 5~MHz bandwidth, which corresponds to
a maximum of 25 Resource Blocks (or 300 subcarriers). The fronthaul was
implemented with a 1~Gbps Ethernet connection among machines with 1400 bytes for
MTU size.

A USRP B210 was connected to the RRU machine to establish radio access to OAI
network. A Samsung Galaxy S4 smartphone was used as UE connecting to RRU through
USRP and generate user plane traffic over FH. Iperf tool was used to create
traffic to UE through OAI network to test and obtain results.

\section{Experimental Results}
\label{sec:results}

\begin{figure}[thb]
\begin{center}
 \includegraphics[width=0.88\columnwidth]{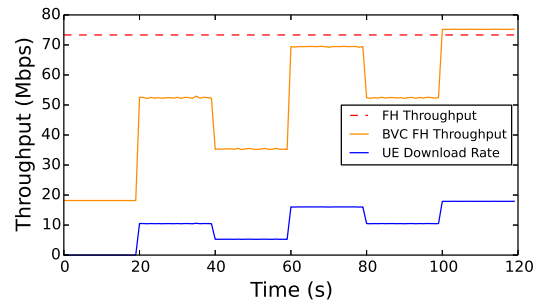}%
 \caption{UE download speed and FH downlink throughput.\label{fig:throughput}}
\end{center}
\end{figure}

The experimental results were captured with the iperf tools, where it was used
an iperf client on UE and an iperf server on EPC machine to generate traffic
over the FH downlink. Fig.~\ref{fig:throughput} shows UE download bit rate and
FH downlink throughput with and without BVC. 

Without BVC, the FH traffic is constant, but when the proposed method is applied
the FH throughput changes with the UE downlink bit rate. As indicated before,
the BVC algorithm compresses the data according to UE download rate. As shown in
Fig.~\ref{fig:throughput}, there is a reduction of 75.75\% on BVC FH throughput
in relation to FH throughput without BVC when there is only UE control data.
Comparing with the FH bitrate necessary to transport time-domain samples
presented in~\cite{chang2017flexcran}, the reduction of the FH bitrate is
approximately 92\%. The BVC FH throughput grows proportionally when the iperf
application starts to require data on the UE. Therefore, with only control data,
the fronthaul usage is about 24.25\% of the total FH capacity, the growth is
about 4.5\% of total FH capacity per 1~MB required by UE.

\begin{figure}[thb]
\begin{center}
 \includegraphics[width=0.8\columnwidth]{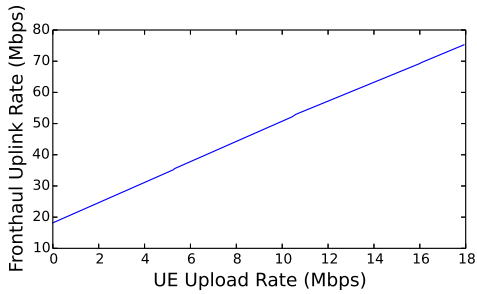}%
 \caption{Relationship between the UE and FH thoughput.\label{fig:throughput2}}
\end{center}
\end{figure}

Actually, there is a linear relation between the UE data rate and the FH
bitrate, as shown in Fig.~\ref{fig:throughput2}. When UE download rate is zero,
there is only control traffic being transported over FH (control channels and
references signals of LTE). After UE starts to require bandwidth, the control
traffic is combined with UE data, which increases FH throughput proportionally
to the number of REs being requested by UE. Therefore, the more bandwidth is
requested by UE, more REs are allocated to transport data, and consequently more
FH rate is needed.

\begin{figure}[thb]
\begin{center}
 \includegraphics[width=0.75\columnwidth]{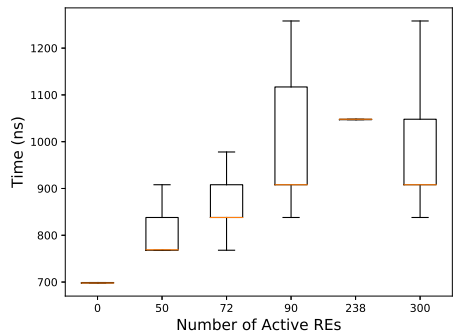}%
 \caption{Time required by compression operations.\label{fig:compression}}
\end{center}
\end{figure}

\begin{figure}[thb]
\begin{center}
 \includegraphics[trim=0 0 0
 40,clip,width=0.75\columnwidth]{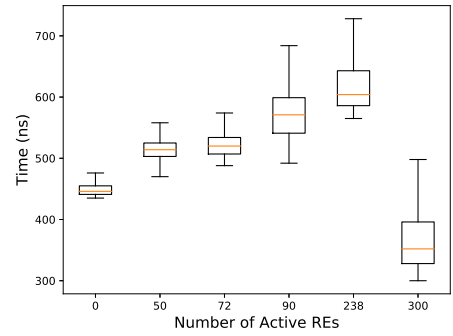}%
 \caption{Time required by decompression operations.}
 \label{fig:decompression}
\end{center}
\end{figure}

The BVC was also evaluated in terms of processing time. Then, the results were
compared with the method shown in~\cite{chang2017flexcran}. where A-law is also
used to reduce FH traffic. Compression and decompression operations time were
collected to measure the processing impact of using the BVC technique.
Considering a 5~MHz scenario (300 REs), it was measured time to execute bit
vector mapping and A-law quantization.
Figures~\ref{fig:compression}~and~\ref{fig:decompression} show boxplots of the
execution time for compression and decompression, respectively. Different values
of active REs were used, where the results of~\cite{chang2017flexcran} are
represented by the case with 300 active REs.

Fig.~\ref{fig:compression} shows the time to compress a certain number of active
REs. It would be expected longer execution time due to the mapping process added
by BVC. However, analyzing the compression time, it was perceived that there is
a trade-off because when a few REs are sent, a few REs will need to be
quantized, which reduces the processing time comparing to the results
of~\cite{chang2017flexcran}(300 REs). It is perceived that up to 90 REs being
used, BVC presents a lower processing time despite the mapping process. With
more than 238 REs, time tends to be longer, but this difference is less than 150
nanoseconds for 238 REs case.

The decompression time is shown in Fig.\ref{fig:decompression}. In this case,
the decompression time for the BVC algorithm is always longer
than~\cite{chang2017flexcran}(300 REs). The results show this behavior because
BVC always need to verify all REs. Therefore, there is no processing gain using
the BVC to decompress LTE symbols on RRU. But processing time overhead is less
than 0.3 microseconds, and the OAI operations were not affected by this
overhead. Moreover, the FH bit rate can be considerably reduced.

\section{Conclusion}
\label{sec:conclusion}

This work has proposed a bit vector mapping method combined with A-law
quantization applied to REs in a packet-based FH. The proposed method works in a
Low-PHY functional split that transports the frequency domain representation of
the signals. The proposed algorithm was implemented and tested in OAI platform
where FH traffic is reduced according to cell load, reaching 75\% reducing when
there is only control traffic for frequency domain, and 92\% of FH bitrate
reduction when compared with the transportation of the time domain signal. The
proposed BVC has presented a low processing overhead in relation to processing
using just A-law quantization, offering a better FH usage. In the future, BVC
can be extended to support uplink.

%

\bibliographystyle{IEEEtran}
\bibliography{zotero_exported_items}

\end{document}